\documentclass[prc,onecolumn,preprint]{revtex4}
\usepackage{color}
\usepackage{dcolumn}
\usepackage{amsmath,amssymb}
\usepackage{graphicx}
\usepackage{bm}
\begin{document}

%\preprint{}
\title{Existence of mesons after deconfinement}
\author{Fabian \surname{Brau}}
\email[E-mail: ]{F.Brau@cwi.nl} 
\affiliation{CWI, P.O. Box 94079, 1090 GB Amsterdam, The Netherlands}
\author{Fabien \surname{Buisseret}}
\email[E-mail: ]{fabien.buisseret@umh.ac.be} 
\affiliation{Groupe de Physique Nucl\'{e}aire Th\'{e}orique,
Universit\'{e} de Mons-Hainaut,
Acad\'{e}mie universitaire Wallonie-Bruxelles,
Place du Parc 20, BE-7000 Mons, Belgium}
\date{\today}

\begin{abstract}
We investigate the possibility for a quark-antiquark pair to form a bound state at temperatures higher than the critical one ($T>T_c$), thus after deconfinement. Our main goal is to find analytical criteria constraining the existence of such mesons. Our formalism relies on a Schr\"{o}dinger equation for which we study the physical consequences of both using the free energy and the internal energy as potential term, assuming a widely accepted temperature-dependent Yukawa form for the free energy and a recently proposed nonperturbative form for the screening mass. We show that using the free energy only allows for the $1S$ bottomonium to be bound above $T_c$, with a dissociation temperature around $1.5\times T_c$. The situation is very different with the internal energy, where we show that no bound states at all can exist in the deconfined phase. But, in this last case, quasi-bound states could be present at higher temperatures because of a positive barrier appearing in the potential. 
\end{abstract}

% Solutions of waves equations: Bound states
% Mesons

\maketitle

\section{Introduction}\label{sec:intro}
The anomalous suppression of heavy meson production in heavy ion collisions has been proposed as a possible signature of deconfinement for about twenty years~\cite{matsui}. The basic idea is that the heavy mesons produced before the formation of a thermalized quark-gluon-plasma would tend to dissociate in the deconfined medium because of the screening of the quark-antiquark ($q\bar q$) interaction by the large number of color charges in the medium. The mechanism is clearly analogous to the Debye screening by electromagnetic charges in QED. Since the pioneering work of Ref.~\cite{matsui}, the suppression of heavy quarkonium production at finite temperature -- that is $T>0$, $T=0$ corresponding to ``usual" QCD -- has been intensively studied in the literature (see Ref.~\cite{intern3} for a review). In particular, the dissociation temperature of a particular meson, that is the temperature at which the quark and the antiquark become unbound, is particularly relevant in order to understand the mechanism of quarkonium dissociation in a quark-gluon-plasma. 

It is a well-known fact that potential models are able to accurately reproduce the experimental meson mass spectra at $T=0$~\cite{old2}. Basically, a meson is then seen as a $q\bar q$ pair interacting via a so-called Cornell potential $ar-4\alpha_s/3r$ (or any other QCD-inspired potential), $r$ being the distance between the quark and the antiquark. The Cornell potential, validated by lattice QCD computations of the energy of a static $q\bar q$ pair~\cite{balirep}, contains a confining linear part and a Coulomb-like term which comes from one-gluon-exchange diagrams. In this potential, $a$ can be interpreted as the tension of a flux tube of length $r$ linking the quark and the antiquark, and $\alpha_s$ is the strong coupling constant. Because of the success of potential models at zero temperature and of the idea that color screening implies modifications of the Cornell interaction, several attempts to understand meson properties at finite temperature within the framework of a Schr\"{o}dinger equation with a phenomenologically modified potential have been made \cite{kar88,rop88,hash,free2}. 

The most accurate way to find a relevant potential term at $T>0$ is again provided by lattice QCD simulations, from which the free energy between a static $q\bar q$ pair at finite temperature can be computed~\cite{mcler,boyd,maez}. A general result of lattice QCD calculations is the observation of a critical temperature, denoted as $T_c$, above which the confining part of the free energy vanishes: That is the deconfinement (see for example Ref.~\cite{kac1}). The {\it free} energy obtained in these calculations, depending both on $T$ and on $r$, can be used as a starting point to build a potential model at $T\neq0$. However, in some other works, it is argued that the \textit{internal} energy should better be used as potential term~\cite{intern1,intern2,intern4}, because of the entropy contribution which is present in the free energy. The situation of how should potential models be applied to $q\bar q$ states at $T>0$ is thus still not completely clarified~\cite{intern4}. 

We propose in the present work to investigate the existence of mesons above the critical temperature by using a Schr\"{o}dinger equation with an appropriate potential term. The motivation to study such a temperature domain does not only come from intrinsic theoretical interest, but also because the current temperatures reached by experiments are in the typical range $(1-2)\times T_c$~\cite{ent}. Let us note that what we call a meson in the following is a $q\bar q$ \textit{bound state}, that is a state with a negative binding energy and an infinite lifetime -- at least formally. Our model, that we present in Sec.~\ref{sec:mod}, is a rather simple one, but it contains the main qualitative features of most of the potential models which were previously developed. Moreover, the Yukawa form that we use for the free energy is in agreement with recent and accurate lattice QCD results~\cite{maez}. In this work, we are mainly interested in qualitative and mostly analytical results constraining the existence of mesons at finite temperature. Actually we propose a simple method which can also be applied to a more complete version of the model studied here. In particular, we focus on the dissociation temperature, which is a physically relevant observable, leading to a direct picture of the evolution of the number of bound states with the temperature. We show that the determination of the dissociation temperature is eventually ruled by a single dimensionless parameter which corresponds to the strength of the potential in a rescaled set of coordinates. Constrains on the dissociation temperature are thus obtained from constrains on the strength of the potential such as it is attractive enough to admit bound states. Since two choices are generally assumed in the literature for the potential term, we study them both in Secs.~\ref{freep} (free energy) and \ref{internp} (internal energy). As we will see in the latter, both possibilities lead to rather different predictions; we then summarize our results in Sec.~\ref{sec:conc}.

\section{The model}\label{sec:mod}

The binding energy of a $q\bar q$ pair in a medium at finite temperature $T>0$ can be obtained by solving the following Schr\"{o}dinger equation
\begin{equation}\label{maineq}
\left[\frac{\vec p^{\, 2}}{2\mu}+V(r,T)\right]	\psi(\vec r,T)= \varepsilon_{n\ell}\, (T)\  \psi(\vec r,T),
\end{equation}
where $\mu=m_qm_{\bar q}/(m_q+m_{\bar q})$ is the reduced mass of the system (we work in units where $\hbar=c=1$). As it is generally done, we assumed a central form for the temperature-dependent interaction potential $V(r,T)$. Consequently, the binding energy $\varepsilon_{n\ell}\, (T)$ does not only depend on the temperature, but also on the radial quantum number $n$ and on the orbital angular momentum $\ell$. An important question is the following: Should the free energy of the $q\bar q$ pair be used as potential term, or should the internal energy be used? Up to our knowledge, the answer is far from being unanimously accepted. Consequently, we will consider both possibilities in this work, and study their consequences on the existence of bound states from the analysis of Eq.~(\ref{maineq}).   

It is known for a long time that lattice calculations can directly compute the free energy, denoted as $F_{\bm 1}(r,T)$, between a quark and an antiquark placed in a thermal bath of gluons and light quarks~\cite{mcler}. For $T\gtrsim T_c$, recent lattice computations agree with a Yukawa form for the free energy of a $q\bar q$ pair in a color singlet, which can be parametrized by~\cite{maez}
\begin{equation}\label{free}
F_{\bm 1}(r,T)=-\frac{4}{3}\, \alpha_s(T)\, \frac{{\rm e}^{-m_D(T) r}}{r},	
\end{equation}
where $\alpha_s(T)$ is an effective strong coupling constant depending on the temperature. Equation~(\ref{free}) shows that the free energy is screened in the finite temperature medium, in analogy to what happens in QED. The potential at $T\gtrsim T_c$ is then no longer a confining one. The screening parameter could be considered as a QCD Debye mass, denoted as $m_D(T)$. Since the free energy naturally emerges from lattice QCD computations, many works directly assumed that $V(r,T)=F_{\bm 1}(r,T)$ to study the properties of mesons with the eigenequation~(\ref{maineq})~\cite{free2,free1}. We point our that, if these works use the free energy as potential term, they do not necessarily take the same expression as ours for $F_{\bm 1}(r,T)$. Actually, the simple form (\ref{free}) has the advantage of being in agreement with the recent results of Ref.~\cite{maez}.

On the other hand, the free energy contains an entropy contribution, and other approaches have suggested that the internal energy, defined as usual in thermodynamics as 
\begin{equation}
U_{\bm 1}(r,T)=	F_{\bm 1}(r,T)+T\, S_{\bm 1}(r,T)= F_{\bm 1}(r,T)+T\, \partial_T F_{\bm 1}(r,T),
\end{equation}
should be preferentially used as potential term in the Schr\"{o}dinger equation~(\ref{maineq}) \cite{intern1,intern2,intern3}. The reason for such a choice could be related to the time scales involved in the system~\cite{intern1}. The first time scale that should be considered is $\tau_b$ is the typical time associated to a particular bound state, i.e. $\left\langle r/\dot r \right\rangle$. The second one is $\tau_h$, that is the time which is needed to transfer heat to matter by changing the entropy $S_{\bm 1}$. If $\tau_b\ll\tau_h$, then the heat transfer can be neglected, and the internal energy is the relevant potential term. If not, the free energy should be used. With the free energy defined as~(\ref{free}), the internal energy reads
\begin{equation}\label{udef}
U_{\bm 1}(r,T)=-\frac{4}{3}\left[\left(T\, \alpha_s(T)\right)'-T\, \alpha_s(T) m_D(T)' r \right]\frac{{\rm e}^{-m_D(T) r}}{r},
\end{equation}
where the prime denotes a derivation with respect to $T$.

The evolution of $F_{\bm 1}(r,T)$ and $U_{\bm 1}(r,T)$ with the temperature is completely fixed by the functions $\alpha_s(T)$ and $m_D(T)$. For the running of the strong coupling constant with the temperature scale, we will assume the well-known one-loop expression~\cite{boyd}
\begin{equation}\label{ast}
	\alpha_s(T)=\frac{2\pi}{\left(11-\frac{2}{3}N_f\right) \ln\left(\frac{T}{\Lambda_\sigma}\right)}.
\end{equation}
From lattice computations, we choose the value 
\begin{equation}\label{lamdef}
\Lambda_\sigma=\beta\, T_c,\ \ {\rm with}\ \ \beta=0.104\pm0.009,
\end{equation}
which has been obtained in Ref.~\cite{boyd}. Actually, $\alpha_s(T)$ is also known at two-loops~\cite{boyd}, but the one-loop formula already captures the essential physical features of the two-loops running coupling constant. Since we are mainly interested in a qualitative description of the existence of mesons versus the temperature, formula~(\ref{ast}) is thus sufficient for our purpose. Lattice QCD also provides a critical temperature $T_c$ appearing in $\alpha_s(T)$. Following the number of light flavors $N_f$, different values can be found, which globally lie in the range $150-300$ MeV~\cite{kac2,bern,kar2}. We will here take the recent estimation of Ref.~\cite{bern}, where $N_f=3$ with two light quarks of the same mass ($u$ and $d$) and one heavier ($s$). It is computed in this work that~\cite{bern}
\begin{equation}\label{tcdef}
T_c=169\pm16\ {\rm MeV}\ \Rightarrow\ \Lambda_\sigma=17.6\pm3.2\ {\rm MeV}.	
\end{equation}

\begin{figure}[ht]
\begin{center}
\includegraphics*[width=9.0cm]{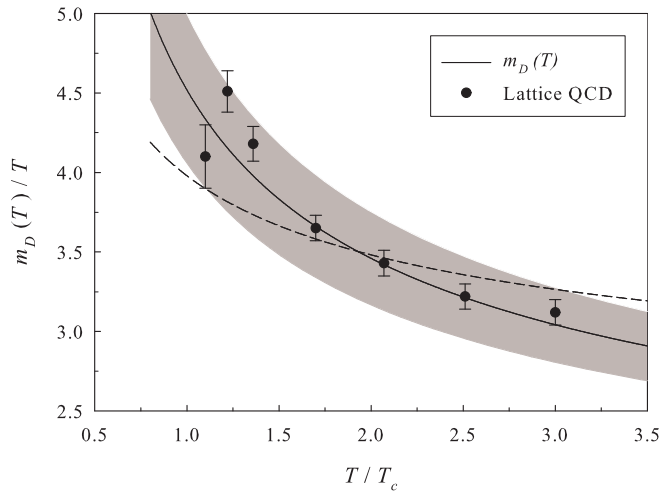}
\end{center}
\caption{Graph of $m_D(T)/T$ as a function of $T/T_c$ for a $q\bar q$ pair in a color singlet. The full circles are taken from the lattice computations of Ref.~\cite{maez}. Formula~(\ref{mdsimo}), computed with the mean values of the parameters, is plotted with a solid line. The error bars on the parameters appearing in $m_D(T)$ actually allow to span the shaded area. The perturbative-like formula~(\ref{mdpert}) is also plotted for comparison with the fitted value $A_f=1.65$ and the mean values of the parameters (dashed line).}
\label{fig1}
\end{figure}

If the function giving $\alpha_s(T)$ is generally accepted, several different forms can be found in the literature for the screening mass $m_D(T)$~\cite{intern2,scr1,scr2,scr3}, depending mainly on the temperature range which is considered. In this paper, we will assume a recently proposed nonperturbative formula, obtained in Ref.~\cite{scr1} within the framework of the background perturbation theory, which states that
\begin{equation}
	m_D(T)=4\pi \eta\, c_\sigma\, \alpha_s(T)\, T,
\end{equation}
with $c_\sigma=0.566\pm0.013$~\cite{boyd} and $\eta=2.06$~\cite{scr1}. Defining $\gamma=4\pi \eta\, c_\sigma$, we have
\begin{equation}\label{mdsimo}
m_D(T)=\gamma\, \alpha_s(T)\, T,\ \  {\rm with}\ \  \gamma=14.652\pm0.337.	
\end{equation}
Contrarily to the well-known form $m_D(T)=[4\pi \left(1+N_f/6\right)]^{1/2} \, [\alpha_s(T)]^{1/2}\, T$~\cite{shur} 
which is assumed to hold at $T\gg T_c$ only~\cite{scr3}, formula~(\ref{mdsimo}) is expected to hold also at $T\gtrsim T_c$. As we show in Fig.~\ref{fig1}, formula~(\ref{mdsimo}) fits indeed rather well the lattice QCD data, taken from Ref.~\cite{maez}, in the range $(1-3)\times\, T_c$. We also tried to fit the data with a perturbative-inspired ansatz of the form~\cite{kac2}
\begin{equation}\label{mdpert}
m_D(T)=A_f\, \sqrt{4\pi \left(1+N_f/6\right)} \, \sqrt{\alpha_s(T)}\ T.	
\end{equation}
Even with the ad hoc value $A_f=1.65$, which leads to the best fit of the data, formula~(\ref{mdpert}) clearly yields a poorer agreement with lattice QCD (see Fig.~\ref{fig1}). Consequently, it is better justified to work with the screening mass defined by Eq.~(\ref{mdsimo}). 

Thanks to Eqs.~(\ref{ast}), (\ref{tcdef}) and (\ref{mdsimo}), the free and internal energies are completely known, and it is possible to compute the binding energies $\varepsilon_{n\ell}\, (T)$ from the eigenequation~(\ref{maineq}). Then, the dissociation temperature of a particular $[(n_0+1),\ell_0]$ state (in spectroscopic notation) can be computed, that is the temperature $T_{n_0\ell_0}$ for which
\begin{equation}
	\varepsilon_{n_0\ell_0}(T_{n_0\ell_0})=0.
\end{equation}
The physical meaning of such a temperature is the following: Once $T\geq T_{n_0\ell_0}$, no $q\bar q$ bound state with quantum numbers $n\geq n_0$ or $\ell\geq\ell_0$ can exist.  

\section{Free energy as potential term}\label{freep}

Let us now investigate the existence of bound states when the free energy~(\ref{free}) is used in the Schr\"{o}dinger equation~(\ref{maineq}): $V(r,T)=F_{\bm 1}(r,T)$. For further computations, it is convenient to work with the dimensionless variables $\vec x,\ \vec q$, defined as
\begin{equation}\label{xqdef}
 \vec x=m_D(T)\ \vec r,\quad \vec q=m_D(T)^{-1}\ \vec p. 
\end{equation}
Equation (\ref{maineq}) can then be rewritten as 
\begin{equation}\label{eqfree}
\left[\vec q^{\, 2}-g\, P(x)\right]	\psi(\vec x,g)= {\cal E}_{n\ell}\, (g)\  \psi(\vec x,g),	
\end{equation}
with
\begin{equation}\label{p_e_def}
	P(x)=\frac{{\rm e}^{-x}}{x},\quad {\cal E}_{n\ell}\, (g)=\left[\frac{2\mu\, \varepsilon_{n\ell}\, (T)}{m_D(T)^2}\right],
\end{equation}
and
\begin{equation}\label{res1}
g=\frac{8\, \mu}{3}\frac{\alpha_s(T)}{m_D(T)}>0.	
\end{equation}
 
Equations~(\ref{eqfree}-\ref{res1}) show that the dimensionless parameter $g$ ruling the depth of the Yukawa potential, $-g\, P(x)$, is the only physical parameter which is responsible for the existence of bound states. In particular, the numbers $g_{n\ell}$ such that ${\cal E}(g_{n\ell})=0$ are the critical strengths of the potential: When $g<g_{n_0\ell_0}$, no bound states with quantum numbers $n\geq n_0$ or $\ell\geq\ell_0$ are present in the spectrum. The numbers $g_{n\ell}$ actually only depend on the form the potential, which is $-g\, {\rm e}^{-x}/x$ in this section. They are given in Table~\ref{tabgnl} for some values of $n$ and $\ell$. The lowest value, namely $g_{00}=1.680$, is such that when $g<g_{00}$, no bound states can form in the potential. We refer the reader to Refs.~\cite{brau1,brau2,brau3,brau4,brau5} for detailed studies on the critical strengths of attractive central potentials, and for techniques to compute them. 
\begin{table}
\caption{Critical strengths $g_{n\ell}$ of the potential $-g\, {\rm e}^{-x}/x$ for some values of $n$ and $\ell$.}
\label{tabgnl}
\begin{center}
\begin{tabular}{c|cc}
\hline\hline
$n/\ell$ & 0 & 1    \\
\hline
0 &  1.680 &  9.082     \\
1 &  6.447  &  17.745     \\
\hline\hline
\end{tabular}
\end{center}
\end{table}

When the free energy is taken as potential term, the dissociation temperature of the $[(n+1),\ell]$ state, denoted as $T_{n\ell}$, is thus given by the solution of equation
\begin{equation}\label{res2}
	m_D(T_{n\ell})=\frac{8\, \mu}{3g_{n\ell}}\alpha_s(T_{n\ell}),
\end{equation}
which is simply a rewriting of the definition~(\ref{res1}).
It is worth mentioning that this result is independent of the explicit form of the strong coupling constant and of the screening mass. The only input, through the values of $g_{n\ell}$, is the Yukawa form~(\ref{free}). Consequently, for another form of the radial potential like $V(r,T)=-a\alpha_s(T)\, m_D(T)\, v[m_D(T)r]$, with $v(x)$ arbitrary, this relation would still be correct [$a=4/3$ in Eq.~(\ref{res2})]; only the value of $g_{n\ell}$ would change. 

If we inject the definition~(\ref{mdsimo}) in (\ref{res2}), we  obtain
\begin{equation}\label{tcdef1}
T_{n\ell}=\frac{8\, \mu}{3\gamma \, g_{n\ell}}.
\end{equation}
Since the model we developed here is physically relevant only above $T_c$, and since $g$ decreases with increasing $T$ [see Eq.~(\ref{res1})], bound states can form only if 
\begin{equation}
	\left.\frac{8\mu\, \alpha_s(T)}{3m_D(T)}\right|_{T=T_c}=\frac{8\, \mu}{3\gamma\, T_c}>g_{00}.
\end{equation}
It implies a lower bound on $\mu$, which reads
\begin{equation}
\label{limitmass}
\mu>\frac{3}{8}g_{00}\gamma\, T_c=1.560\pm0.036\, \ {\rm GeV}.	
\end{equation}
Following the values of the Particle Data Book concerning the current quark masses~\cite{PDG}, this lower bound only allows for  $b\bar b$ bound states -- for completeness, we recall indeed that the most recent experimental values are $m_c=1.25\pm0.09$ GeV and $m_b=4.45\pm0.32$ GeV~\cite{PDG}. Let us note that, in a Schr\"{o}dinger-based formalism, constituent quark masses are more commonly used. The intuitive idea is that a confined quark acquires an additional mass in a hadron because of the particles-antiparticles pairs which are created around it. The difference is particularly important for light quarks: Although the current mass of the $u$ or $d$ quarks is nearly zero, their constituent mass is often taken to be around $0.3$ GeV~\cite{Lucha}. The use of the constituent mass then allows to deal with a well-defined kinetic operator $\vec p^{\, 2}/2m$. For heavy quarks, the current and constituent masses are essentially the same, since the current mass is much greater than the additional mass coming from any particles-antiparticles cloud. Nevertheless, the constraint~(\ref{limitmass}) forbids the existence of mesons made of at least one light quark above $T_c$, either the current or the constituent quark mass is used.

Following Eq.~(\ref{tcdef1}), the highest dissociation temperature is $T_{00}$ for a $b\bar b$ pair, thus the dissociation temperature of the $1S$ bottomonium. It is readily computed that, in this case, $T_{00}=(1.426\pm 0.070)\times T_c$. The other values for $T_{n\ell}$ are always smaller than $T_c$ even for the $b\bar b$ states, thus outside the validity range of our model. Interestingly, the prediction that even the heaviest mesons should be dissociated around $1.5\times T_c$ is in qualitative agreement with a recent work based on the study of quarkonium correlators and spectral functions at nonzero temperature~\cite{pet}. Let us note that, since our model does not take the spin interactions into account, we only have a single dissociation temperature for the both the $\eta_b(1S)$ and $\Upsilon(1S)$ mesons. But, as it can be checked in Refs.~\cite{free2,intern1}, spin-dependent interactions are expected to affect the dissociation temperatures by less than $10\, \%$.

\section{Internal energy as potential term}\label{internp}

\begin{figure}[ht]
\begin{center}
\includegraphics*[width=9.0cm]{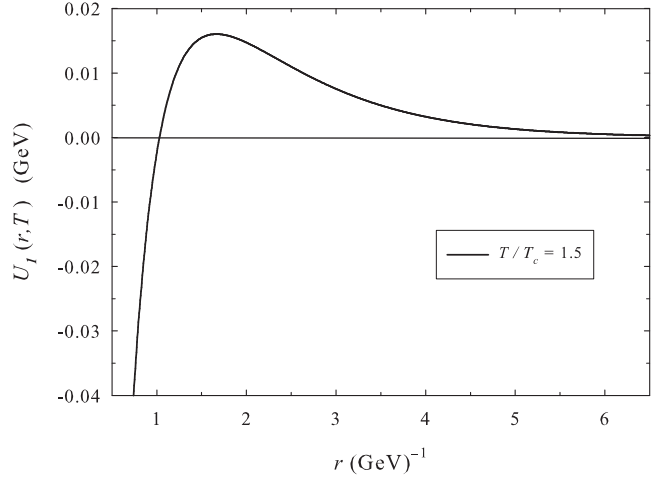}
\end{center}
\caption{Plot of $U_{\bm 1}(r,T)$ computed for $T=1.5\times T_c$ and with the optimal values of the parameters.}
\label{fig3}
\end{figure}

We turn now our attention to the case where the internal energy is taken as potential term in the eigenvalue problem to solve. Since, following Eq.~(\ref{mdsimo}), the screening mass is proportional to $T\, \alpha_s(T)$, the internal energy~(\ref{udef}) can be rewritten as
\begin{equation}\label{freedef}
V(r,T)=U_{\bm 1}(r,T)=-\frac{4}{3}\, \left[T\, \alpha_s(T)\right]'\left[1-m_D(T)r\right]\frac{e^{-m_D(T) r}}{r}.	
\end{equation}
The sign of the internal energy crucially on the one of $\left[T\, \alpha_s(T)\right]'$. It is easily checked from definition~(\ref{ast}) that, for $T>T_c$, $\left[T\, \alpha_s(T)\right]'>0$. As an illustration, we plotted $U_{\bm 1}(r,T)$ for $T=1.5\times T_c$ in Fig.~\ref{fig3}.
    
Let us begin by a study of the possible bound states. The main equation~(\ref{maineq}) can again be rewritten in the form of Eq.~(\ref{eqfree}), but with 
\begin{equation}\label{meta}
	P(x)=(1-x)\frac{{\rm e}^{-x}}{x},
\end{equation}
and $g>0$ defined as
\begin{equation}\label{res3}
	g=\frac{8\, \mu}{3}\frac{\left[T\, \alpha_s(T)\right]'}{m_D(T)}=\frac{8\, \mu}{3\gamma}\left[\frac{1}{T}+\frac{\alpha_s(T)'}{\alpha_s(T)}\right].
\end{equation}

The critical strength $g_{00}$ is equal to $4.937$ for the potential (\ref{meta}). We still refer the reader to Refs.~\cite{brau1,brau2,brau3,brau4,brau5} for methods to compute $g_{00}$. Consequently, bound states will exist if 
\begin{equation}\label{crit1}
	\frac{8\, \mu}{3\gamma}\left[\frac{1}{T}+\frac{\alpha_s(T)'}{\alpha_s(T)}\right]=\frac{8\, \mu}{3\gamma\,  T}\left[1-\frac{1}{\ln\left(T/\Lambda_\sigma\right)}\right]>g_{00}.
\end{equation}
Since the function of $T$ in this last relation is ever decreasing for $T\geq T_c$, a lower bound on $\mu$ is obtained when Eq.~(\ref{crit1}) is computed for $T=T_c$. One finds
\begin{equation}
	\mu>\frac{3\gamma\, T_c}{8}\, g_{00}\, \left[\frac{\ln\beta}{\ln\beta+1}\right]= 8.213\pm 1.215\ {\rm GeV}.
\end{equation}
Apart from hypothetical bound states involving top quarks, this condition states that no bound state can survive after deconfinement provided the internal energy is used as potential term. The minimal reduced mass is indeed higher than any physically allowed reduced mass. 

\begin{figure}[ht]
\begin{center}
\includegraphics*[width=9.0cm]{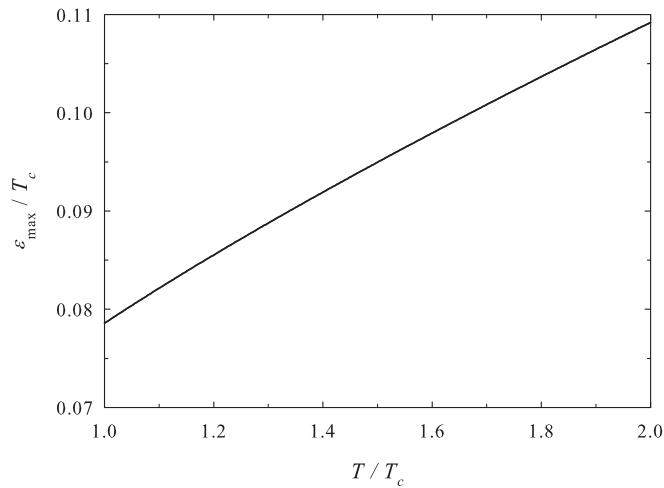}
\end{center}
\caption{Plot of the maximal energy $\varepsilon_{{\rm max}}(T)$ for the existence of quasi-bound states states as a function of the temperature in units of $T_c$.}
\label{fig5}
\end{figure}

If bound states cannot form at high temperatures, it should be however possible to find quasi-bound states in the potential $U_{{\bm 1}}(r,T)$, i.e. resonance states with a positive eigenenergy and a finite lifetime. This is due to the presence of a positive barrier in $U_{{\bm 1}}(r,T)$, as it can be observed in Fig.~\ref{fig3}. The function $P(x)$ given by relation~(\ref{meta}) has a minimum in $x=M$, with $M=(1+\sqrt 5)/2$. The maximum value of the potential $-g P(x)$ is thus reached in $x=M$, and quasi-bound states are expected to appear for 
\begin{equation}\label{ineq}
{\cal E}<-gP(M).	
\end{equation}
These states will have a lifetime which goes smaller as ${\cal E}$ gets closer of $-gP(M)$. We stress that the results we discuss here concerning quasi-bound states are only valid in the $s$-channel. A rewriting of inequality~(\ref{ineq}) leads to an upper bound on the physical binding energy, that is 
\begin{equation}
\varepsilon<\varepsilon_{{\rm max}}(T)=\frac{4}{3}(M-1)\frac{{\rm e}^{-M}}{M}\, m_D(T)\, [\alpha_s(T)\, T]'.	
\end{equation}
Figure~\ref{fig5} shows the evolution of the upper bound $\varepsilon_{{\rm max}}(T)$ with the temperature. Since it is always increasing, the formation of quasi-bound states seems favored at high temperatures. However, one can expect that, for increasing $T$, the average energy $\varepsilon$ of a $q\bar q$ pair in the deconfined medium will increase. Consequently, even if the particular shape of the internal energy gives more chance for the existence of quasi-bound states at high temperature, an ever decreasing number of $q\bar q$ pairs will have the appropriate energy $\varepsilon$. A detailed study of these quasi-bound states thus requires a careful analysis involving statistical mechanical arguments, which is out of the scope of this paper. Finally, we can notice that the radius of these states should necessarily be smaller than $x=M$. It implies that their physical radius is such that $r<M/m_D(T)$. This upper bound tends to zero as $T$ becomes larger. 

\section{Summary of the results}\label{sec:conc}

In this work, we have investigated the possible existence of mesons beyond the critical temperature. We have built a potential model from a commonly accepted Yukawa form for the free energy, that is Eq.~(\ref{free}). We used the well-known one-loop expression for the running coupling constant which is simpler than the two-loops expression and contains mainly the same physical information. For the screening mass however, we took the nonperturbative result of Ref.~\cite{scr1}, which predicts that $m_D(T)\propto \alpha_s(T)\, T$ instead of the usual dependence in $\sqrt{\alpha_s(T)}\, T$. We have checked that the expression~(\ref{mdsimo}) that we used for $m_D(T)$ is indeed in better agreement with recent lattice QCD simulations~\cite{maez} than the perturbative-like form~(\ref{mdpert}) in the temperature range $(1-3)\times\, T_c$. 

We provide in this paper a simple method, which allows us to get mainly analytical results and can be applied to different versions of the model discussed here. In particular, we have shown that the determination of the dissociation temperature is ruled by a single dimensionless parameter which corresponds to the strength of the potential in a rescaled set of coordinates. Constrains on the dissociation temperature are thus obtained from constrains on the strength of the potential such as it is attractive enough to admit bound states. We observed very different behaviors following that the free energy or the internal energy is chosen as potential term in the Schr\"{o}dinger equation we study.

When the free energy is used, we have shown that the dissociation temperature of a given meson, that is the temperature at which it becomes unbound, can be analytically obtained. Logically, the dissociation temperature {\it decreases} for {\it increasing} $n$ and $\ell$, but also for {\it decreasing} $\mu$. We deduced from Eq.~(\ref{tcdef1}) that the only state which could be present above $T_c$ is the $1S$ bottomonium, with a dissociation temperature of $(1.426\pm 0.070)\times T_c$, in qualitative agreement with another recent work~\cite{pet}. 

Using the internal energy as potential term leads to a rather different picture. The existence of bound states is still only ruled by a dimensionless parameter [see Eq.~(\ref{res3})], but we have shown that ``true mesons", i.e. $q\bar q$ bound states, cannot exist anymore above $T_c$ in this case. However, it is important to notice that the potential now exhibits a positive barrier even at zero angular momentum. This allows for the existence of quasi-bound states (metastable states) in the continuum, i.e. for positive eigenenergy. In this picture, a few heavy $q\bar q$ pairs could thus subsist in the medium above $T_c$ as compact (small radius) quasi-bound states.

\acknowledgements
The authors thank Dr. Francis Michel for constant interest in this work, and for useful advices. F. Buisseret thanks the FNRS Belgium for financial support. F. Brau acknowledges support by the ``Dynamics of Patterns'' program of the Netherlands Organization for Scientific research NWO.

\end{document}